\documentclass[twoside,twocolumn,10pt]{article}

\usepackage[T1]{fontenc}
\usepackage[utf8]{inputenc}
\usepackage[english]{babel}
\usepackage{microtype}

\usepackage[
  left=1.7cm,
  right=1.7cm,
  top=2.0cm,
  bottom=2.0cm,
  columnsep=7mm
]{geometry}
\usepackage{lmodern} % police claire et standard
\usepackage{setspace}
\usepackage{amsmath,amssymb,amsfonts,bm,mathtools}
\usepackage{booktabs}
\usepackage{ulem}
\usepackage[version=3]{mhchem}
\usepackage[
  locale=US,
  separate-uncertainty,
  per-mode=symbol,
  range-phrase={ to },
  range-units=single,
  list-units = single
]{siunitx}
\DeclareSIUnit{\OD}{\text{OD}}
\DeclareSIUnit{\uM}{\text{µM}}
\DeclareSIUnit{\mM}{\text{mM}}
\DeclareSIUnit{\M}{\text{M}}
\DeclareSIUnit{\nL}{\text{nL}}
\DeclareSIUnit{\Angst}{\text{Å}}

\usepackage{graphicx}
\usepackage{subcaption}
\usepackage[font=small,labelfont=bf]{caption}
\usepackage{float}
\usepackage{pifont}

\usepackage[dvipsnames]{xcolor}
\usepackage[
  colorlinks=true,
  linkcolor=MidnightBlue,
  citecolor=ForestGreen,
  urlcolor=BrickRed
]{hyperref}

\usepackage[sort&compress,numbers]{natbib}
\usepackage{titlesec}
\titleformat{\section}{\large\bfseries\sffamily}{\thesection}{0.5em}{}
\titleformat{\subsection}{\normalsize\bfseries\sffamily}{\thesubsection}{0.5em}{}
\titlespacing*{\section}{0pt}{1ex}{0.5ex}
\titlespacing*{\subsection}{0pt}{0.5ex}{0.3ex}

\usepackage{fancyhdr}
\definecolor{cream}{RGB}{222,217,201}
\definecolor{spect0}{RGB}{158,1,100}
\definecolor{spect5}{RGB}{244,109,67}
\definecolor{spect25}{RGB}{94,79,162}
\definecolor{cred}{RGB}{196,78,82}

\def\chanone{\raisebox{.5pt}{\textcircled{\raisebox{-.9pt}{1}}}}
\def\chantwo{\raisebox{.5pt}{\textcircled{\raisebox{-.9pt}{2}}}}
\def\chanthree{\raisebox{.5pt}{\textcircled{\raisebox{-.9pt}{3}}}}
\def\chani{\raisebox{.5pt}{\textcircled{\raisebox{-.9pt}{i}}}}

\newcommand\titlediffhelper{Local chemotactic response of \textit{Escherichia coli} in fluid and near surfaces}
\newcommand\abstractdiffhelper{Bacteria can adjust their swimming behaviour in response to chemical variations, a phenomenon known as chemotaxis. This process is characterised by a drift velocity that depends non-linearly on the concentration of chemical species and its "local" gradient. To study this process more effectively, we optimised a 3-channel microfluidic device to generate a stable, linear concentration profile of chemoattractants. This setup allows us to monitor the response of \textit{Escherichia coli} to casamino acids or $\alpha$-methyl-DL-aspartic acid at the individual level. By analysing the movement of a population of individuals both in fluid and on surfaces, we achieve faster, more accurate quantification of the population's chemotactic response. In the fluid, the chemotactic response is described by the equation $v_c=\chi(c) \nabla c$, with $\chi(c) = \chi_0 /[(1 + c/c_-)(1 + c/c_+)]$ the chemotactic susceptibility. For $c_- \ll c \ll c_+$, i.e. when bacteria perform chemotaxis, the bacterial chemotactic velocity is proportional to the concentration gradient divided by the concentration and $v_c \propto \nabla c/c = \nabla (\log c)$. However, on surfaces, the chemotactic flux is inhibited.}

\title{\textbf{\titlediffhelper}}

\author{
    Adam Gargasson$^{1}$, Julien Bouvard$^{1}$, Carine Douarche$^{1}$,\\
    Peter Mergaert$^{2}$, Harold Auradou$^{1}$\\[4pt]
    \small $^{1}$Université Paris-Saclay, CNRS, FAST, 91405 Orsay, France\\
    \small $^{2}$Université Paris-Saclay, CNRS, I2BC, 91190 Gif-sur-Yvette, France\\[3pt]
    \small Correspondence: \href{mailto:harold.auradou@universite-paris-saclay.fr}{harold.auradou@universite-paris-saclay.fr}
}

\date{\today}

\begin{document}
%%%%%%%%%%%%%%%%%%%%%%%%%%%%%%%%%%%%%%%%%%%%%%%%%%%%%%%%%%%%

\twocolumn[
\begin{@twocolumnfalse}
    \maketitle
    \vspace{-0.5em}
    \begin{abstract}
       \noindent
       \abstractdiffhelper
    \end{abstract}
    \vspace{1em}
\end{@twocolumnfalse}
]

%%%%%%%%%%%%%%%%%%%%%%%%%%%%%%%%%%%%%%%%%%%%%%%%%%%%%%%%%%%%
% MAIN TEXT
%%%%%%%%%%%%%%%%%%%%%%%%%%%%%%%%%%%%%%%%%%%%%%%%%%%%%%%%%%%%

\section{Introduction}

Chemotaxis is a phenomenon that allows bacteria to actively modulate their movement to move up or down chemical gradients. Bacteria are then able to detect and navigate towards nutrients while avoiding potential toxins or adverse conditions~\cite{Berg_1975,Adler_1975,Karmakar_2021}. Chemotaxis is an essential phenomenon in many biological processes, including microbiome assembly~\cite{Raina2022}, symbiotic interactions~\cite{raina2019role} such as plant root colonisation~\cite{Tsai2025} or insect gut infection~\cite{Lextrait2025}, as well as biofilm formation~\cite{Merritt_2007,armitano2013aerotaxis,stricker2020hybrid,Keegstra2022}. It also allows bacteria to move towards polluted areas of soil or sediments and is expected to contribute to the optimisation of soil bio-decontamination technologies~\cite{Ford_Harvey_2007,Bhushan_Samanta_Chauhan_Chakraborti_Jain_2000,Debarati_2006,Harms_2006,Yamamoto-Tamura_2015}.

Various methods are available to study bacterial chemotaxis systematically: capillary assays~\cite{ADLER1973}, densitometry assays~\cite{dahlquist1972quantitative,holz1978quasi}, stopped-flow diffusion chambers~\cite{Ford_1991,Lewus_2001}, or swarm plate assays~\cite{MoralesSoto2015,Cremer2019}. These traditional methods are accessible and well-established; however, the chemical gradients they create are challenging to control and can change over time. The use of microfluidic technologies in microchip development addresses this challenge by enabling the creation of controlled environments on micro- to millimetre scales. Microfluidic chips enable the visualisation of bacterial behaviour using microscopy. This capability has facilitated the reconstruction of bacterial trajectories and the assessment of their motility in a specific environment. This technology has advanced quickly, leading to the development of various systems that enable the observation of microorganisms in different gradients~\cite{Kim_Kim_Jeon_2010,Ahmed_2010,Wu2013,Grognot_2021,Yue2026}. 

\begin{figure*}[htpb]
    \centering
    \includegraphics[width = \linewidth]{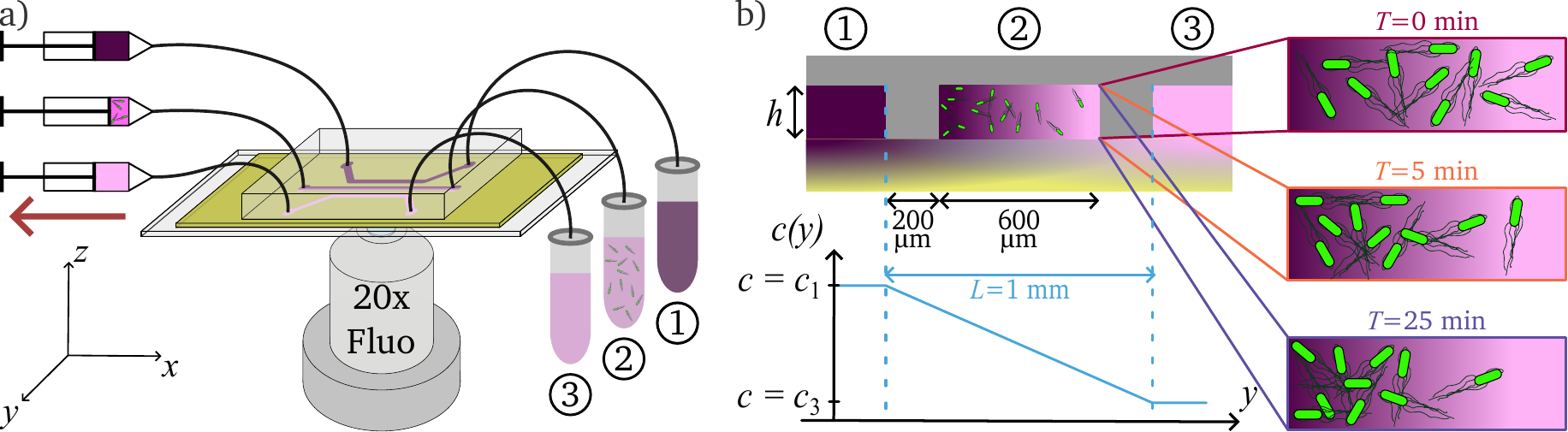}
    \caption{\textbf{Experimental setup.} (a) The microfluidic chip, placed on an inverted microscope and observed under $20 \times$ magnification, consists of three parallel channels. The channels \chanone~and \chanthree~are filled with a chemoattractant at concentrations $c_1$ and $c_3$ diluted in the motility buffer (MB). The reference frame drawn will be used throughout the whole article. (b) The chemoattractant diffuses through the agar layer, creating a uniform gradient between channels \chanone~and \chanthree, notably in channel \chantwo~where fluorescent \textit{E. coli} bacteria are injected. At $T = 0$, the bacteria are homogeneously distributed along the $y$-axis before starting to accumulate towards the chemoattractant source. In our experiments, we selected the axes to ensure that the highest concentration is in channel \chanone, i.e., $c_1 > c_3$.}
    \label{fig:setup}
\end{figure*}

One of these methods is based on the use of a three-channel microfluidic chip~\cite{Diao_2006,Cheng_2007,Kalinin_Jiang_Tu_Wu_2009,Ahmed_Shimizu_Stocker_2010,Son2016} (see Fig.~\ref{fig:setup}), composed of three parallel channels between which chemical species can diffuse. 
This setup maintains a steady, uniform chemical gradient by continuously flowing solutions of different concentrations through the two outermost channels. This approach prevents the time-dependent weakening of the gradient caused by molecular diffusion in existing devices, thereby enabling more precise characterisation of chemotaxis.
The bacteria are placed in the central channel and observed by a microscope. Chemotaxis can then be directly observed via the population migration along the gradient. Several setups using a 3-channel chip have been used to quantify the chemotactic or aerotactic behaviour of bacteria~\cite{Ahmed_2010,Menolascina_Rusconi_Fernandez_Smriga_Aminzare_Sontag_Stocker_2017,morse2016aerotactic}, particularly focusing on \textit{E. coli} chemotaxis toward $\alpha$-methyl-DL-aspartic acid (MeAsp)~\cite{Kalinin_Jiang_Tu_Wu_2009,Ahmed_Shimizu_Stocker_2010,Gargasson2025}. In these studies, chemotaxis was characterised by observing the stationary distribution of bacteria across the width of the central channel.
This steady-state regime is typically achieved after several minutes of bacterial migration along the gradient. In this paper, we propose a method that eliminates the need for this condition. Our approach enables the determination of the chemotactic velocity from trajectories derived from time-lapse images acquired at any time during the experiment.

To comprehend the proposed method, we need to start from the convection-diffusion equation, which describes the evolution of the density of a bacterial population $b(\mathbf{r},t)$~\cite{keller1971model,Rivero_1989}: 
\begin{equation}
    \dfrac{\partial b}{\partial t} + \boldsymbol{\nabla}\cdot\mathbf{J} = 0, \qquad \mathbf{J} = b \mathbf{v_c} - \mu(c) \boldsymbol{\nabla}{b}=b\mathbf{v},
    \label{eq:ks}
\end{equation} 
with $\mathbf{J}$ the bacterial flux in number of bacteria crossing a surface per unit of time, and $c$ the concentration of chemoattractant. The first contribution in $\mathbf{J}$ is the chemotactic flux, i.e. the bacterial flow caused by the inhomogeneity of the chemoattractant distribution. The second contribution in Eq.~\eqref{eq:ks} is the diffusive flux, with $\mu(c)$ the diffusion coefficient associated with random swimming. The net flux resulting from the two fluxes is $b\mathbf{v}$. In the stationary phase, the chemotactic velocity is simply balanced by diffusion, resulting in a bacterial concentration profile decreasing exponentially. The chemotactic velocity $v_c$ is then calculated using the characteristic length $\lambda$ associated with the exponential decrease of the profile $b(y)$ as $v_c = \mu / \lambda$. We propose an alternative method that measures both the net and diffusive fluxes, which can be determined from the positions and trajectories of bacteria analysed using video techniques. In theory, the chemotactic velocity can be derived from these two quantities using Eq.~(\ref{eq:ks}). We will test the conditions under which the method's assumptions remain valid and determine the limits of detection.

In practice, the key question is to determine the relationship between chemotactic velocity, chemoattractant concentration, and its gradient. 
%A great amount of experimental and modelling effort has been directed at determining the dependence of the chemotactic velocity on the attractant concentration $c$. 
In the limit of shallow gradients (i.e. $v_c/v_s \ll 1$ where $v_s$ is the swimming velocity)~\cite{Segel_1977,Rivero_1989} along the direction $y$, the chemotactic velocity $v_c$ is proportional to the attractant gradient with 
\begin{equation}
    v_c = \chi(c) \nabla c%\partial c/\partial y,
    \label{chi(c)}
\end{equation}
where $\chi(c)$ represents the chemotactic coefficient, accounting for the non-linear dependence of the chemotactic velocity on the attractant concentration.~\cite{Keller_Segel_1971,Rivero_1989,Ford_1991,Lewus_2001}. Recent studies revealed the complexity of the time scales associated with the cellular signalling pathway~\cite{Hazelbauer_Falke_Parkinson_2008}. They showed the existence of an adaptation dynamics of the signal transduction pathway, mediated by the chemosensors' methylation, which is slow compared to the typical run time, revealing the need to include these dynamics in the coarse-grained models~\cite{Laganenka_Colin_Sourjik_2016}. Thus, additional variables like the average methylation level of the receptors~\cite{Tu_Shimizu_Berg_2008,Kalinin_Jiang_Tu_Wu_2009,Si_Wu_Ouyang_Tu_2012} need to be considered. When the stimulus variation is relatively small, the model gives
\begin{equation}
    \chi(c)= \frac{\chi_0}{(1 + c/c_-)(1 + c/c_+)}
    \label{eq:SiTu}
\end{equation}
where $c_-$ and $c_+$ are the binding affinities of the receptor at respectively low and high methylation level~\cite{Si_Wu_Ouyang_Tu_2012,Colin_Zhang_Wilson_2014}. For $c_-\ll c\ll c_+$, Eq.~\eqref{eq:SiTu} gives $\chi(c)\propto 1/c$ or $v_c \propto \nabla c/c \sim \ln(\nabla c)$~\cite{Kalinin_Jiang_Tu_Wu_2009,Menolascina_Rusconi_Fernandez_Smriga_Aminzare_Sontag_Stocker_2017}. This susceptibility is not always observed for aerotaxis~\cite{Bibikov2004}. For some bacteria such as \textit{Shewanella oneidensis} and \textit{Burkholderia contaminans}, $v_c \propto \nabla c / c^2$ is observed~\cite{Stricker2020,Bouvard_2022}. Using our method applied to experiments conducted under various gradient conditions, we demonstrate that the chemotactic susceptibility of an \textit{E. coli} strain follows the relationship proposed by Eq.~(\ref{eq:SiTu}). By comparing our method with the stationary method, we will furthermore show that the proposed approach extends the range of chemoattractant concentrations over which a chemotactic response can be detected. We will also show that the method can be applied "locally", i.e., in specific regions in the canal. This will be done by segmenting the images into stripes normal to the gradient and performing the analysis in each stripe. 

Finally, the method will be applied to time-lapse images acquired at different distances from surfaces. This will highlight the significant influence of surfaces on bacterial chemotaxis, with important implications for understanding this process in porous media. 
%To achieve our goal, we conduct measurements of bacterial trajectories to quantify the chemotactic velocity $v_c(c)$ and susceptibility $\chi(c)$ of a population of \textit{E. coli} bacteria in response to various gradients of MeAsp and casamino acids.

\begin{table}
    \begin{center}
        \begin{tabular}{rp{0.85\linewidth}}
            \toprule
            $c_i$ & Chemoattractant concentration in channel \chani\\
            $\nabla c$ & Concentration gradient between channels \chanone~and \chanthree\\
            $\bar{c}$ & Average concentration in channel \chantwo\\
            $\Delta y$ & Stripe width over which trajectories' characteristics are averaged\\
            $L$ & Distance between channels \chanone~and \chanthree\\
            $h$ & Channels height\\
            $T$ & Time elapsed since the beginning of each experiment\\
            $t$ & Time since the acquisition of a time-lapse image\\
            $b$ & Bacterial density\\
            $\mathbf{J}$ & Bacterial flux\\
            $\tau_c$ & Velocity correlation time of the bacteria (pop. average)\\
            $l_c$ & Velocity correlation length of the bacteria (pop. average)\\
            $\mu$ & Diffusion coefficient of the bacteria (pop. average)\\
            $v_s$ & Average swimming velocity of the bacteria (pop. average)\\
            $v$ & Net velocity (pop. average)\\
            $v_{\mu}$ & Diffusive velocity (pop. average)\\
            $v_c$ & Chemotactic velocity (pop. average)\\
            $\tilde{v}_{...}$ & $= v_{...} / v_s$ Normalised velocity (pop. average)\\
            $\chi$ & Chemotactic susceptibility (population average)\\
            $\lambda$ & Characteristic length of the exponential decrease in the stationary density profile of accumulated bacteria\\
            $D_\mathrm{j}^\mathrm{i}$ & Brownian diffusion coefficient of a solute i in a media j\\
            \bottomrule
        \end{tabular}
    \end{center}
    \caption{Notations and grandeurs used in this article.}
    \label{tab:nota}
\end{table}

\section{Materials and methods}

\subsection{Microorganism and Growth Conditions}

\textit{Escherichia coli} strain RP437-YFP expressing yellow fluorescent protein was grown in a minimal salt medium (M9G), consisting of M9 minimal medium salts supplemented with \qty{0.1}{\percent} casamino acids, \qty{0.4}{\percent} glucose, \qty{0.1}{\mM} calcium chloride and \qty{2}{\mM} magnesium sulfate in milliQ water. To this growth medium, \qty{25}{\ug\per\mL} chloramphenicol (CAM) was added to select for the plasmid carrying the yellow fluorescent protein gene. The bacteria were cultured in an incubator shaker at \qty{240}{rpm} and \qty{30}{\celsius}. A preculture was grown to an OD close to \num{0.3} and washed twice by centrifugation (\qty{2300}{g} for \qty{10}{min}). The bacteria were then re-suspended into a minimal motility buffer (MB; \qty{0.1}{\M} EDTA, \qty{1}{\mM} methionine, \qty{1}{\M} sodium lactate, and \qty{0.1}{\M} Phosphate buffer in milliQ water at pH = \num{7.0}). This medium provides the salts needed for bacterial swimming, but without any nutrients to prevent bacterial growth. The final suspension concentration is initially adjusted to an optical density (OD) between \numlist{0.06;0.1}.

\subsection{Fabrication and assembly of the setup}

The protocol follows the work of Gargasson \textit{et al.}~\cite{Gargasson2025}. In brief, we use a microfluidic chip constructed with three parallel channels, each \qty{600}{\um} wide, separated by walls that are \qty{200}{\um} thick, and with a height $h = \qty{120}{\um}$ (see Fig.~\ref{fig:setup}). The chip is made from polydimethylsiloxane (PDMS), which allows oxygen to permeate, ensuring consistent bacterial motility throughout the experiments. This chip is positioned on a \qty{1}{\mm} thick layer of a \qty{3}{\percent} agar gel, which is placed on a glass slide. Chemical solutions and bacterial suspensions are withdrawn into the device at a rate of \qty{10}{\uL\per\minute} using a syringe pump. Special care is taken during setup and installation to prevent air bubbles from entering the channels and tubes.

\subsection{Chemotaxis experiment}
\label{sec:chexp}

The experiment begins by placing the drain and source inlet tubes in microtubes containing the chemoattractants at the concentrations to be tested. Once the fluids have reached the channels, \textit{i.e} after \qty{10}{\minute}, the flow rate is reduced to \qty{1}{\uL\per\minute} in channels \chanone~and \chanthree, and to \qty{0}{\uL\per\minute} in channel \chantwo. \qty{30}{\minute} are to be waited for a uniform, steady gradient to be established (see SM~\ref{sec:steady})~\cite{Gargasson2025}. When the steady gradient is reached, the central channel pump is temporarily reactivated for \qty{30}{\s} to fully renew the bacterial population without altering the gradient. Then, the valve connecting the inlet to the microtube containing the bacterial suspension is closed to reduce any residual flux along the channel \chantwo. Note that the flows in channels \chanone~and \chanthree~are kept constant at \qty{1}{\uL\per\minute} throughout the entire duration of the experiment, to maintain a steady chemical gradient in channel \chantwo.

The central channel is finally observed under fluorescence with a $\num{20} \times$ objective. The field of view is $\qty{660}{\um} \times \qty{660}{\um}$ to allow for the imaging of the entire width of the channel, with a field depth of \qty{37(5)}{\um}.

We used two chemoattractants: casamino acids, obtained from MP Biomedicals LLC, and $\alpha$-methyl-DL-aspartic acid (MeAsp), obtained from MedChemExpress. Casamino acids are a mixture of amino acids and its molecular weight was taken to be approximately \qty{100}{\g\per\mol}, while $\alpha$-methyl-DL-aspartic acid is \ce{C5H9NO4} with a molecular weight of \qty{147}{\g\per\mol}. The average concentration is defined as $\bar{c} = (c_1 + c_3) / 2$ and the gradient is $\nabla c = (c_3 - c_1) / L$. Some experiments are performed with $c_3 = \qty{0}{\mM}$; in this case, the value of $\nabla c / \bar{c} = -2/L$ is constant. In all our experiments, $\nabla c < 0$ because we define the axes such that the channel with the highest concentration corresponds to channel \chanone. Consequently, if the bacteria are attracted to the chemoattractant, their chemotactic velocity $v_c$ will be negative.

\subsection{Data analysis of the bacteria trajectories}
\label{sec:analyse}

\begin{figure*}[htpb]
    \centering
    \includegraphics[width = \linewidth]{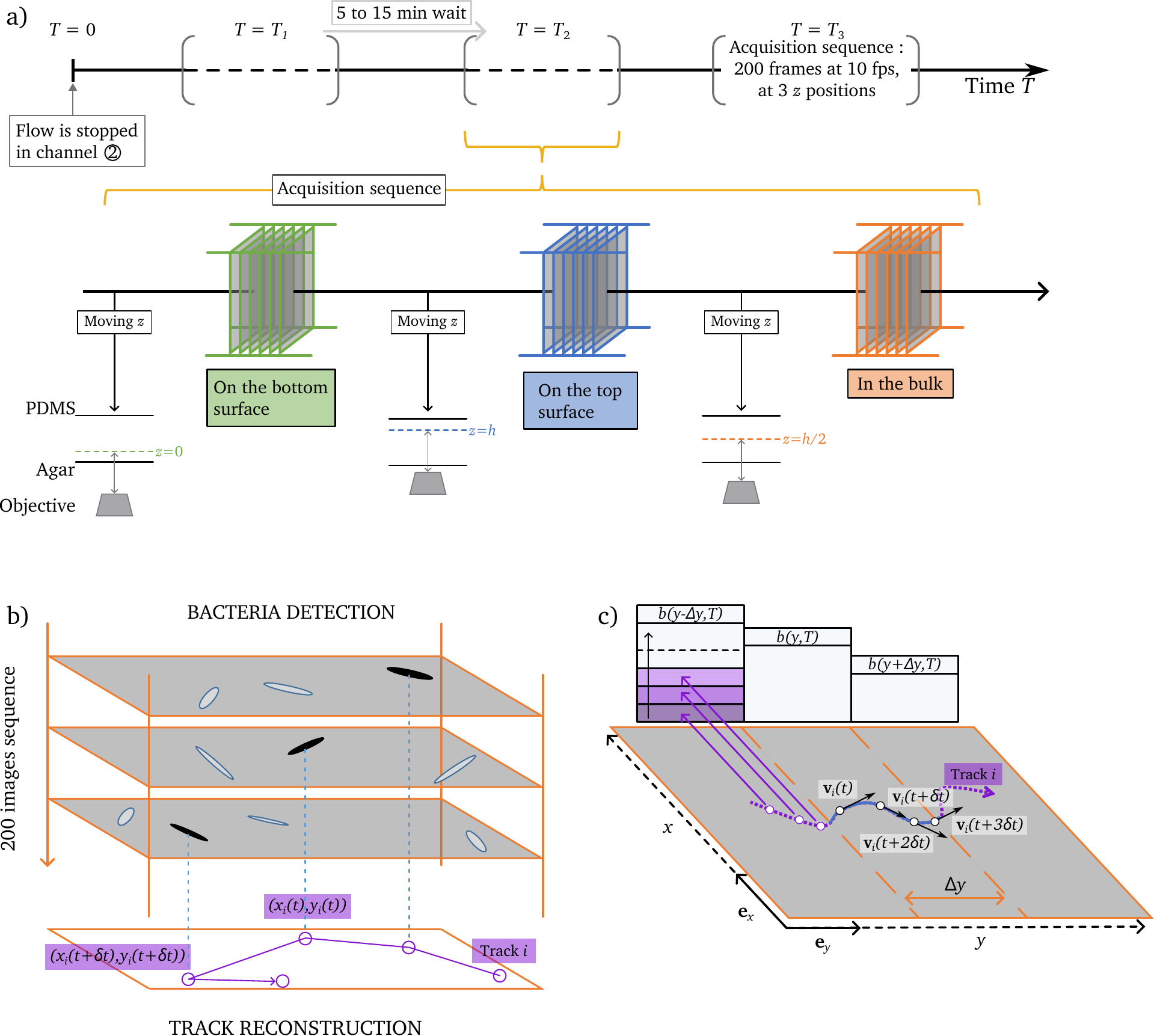}
    \caption{\textbf{Illustration of the three key steps of our experiments:} (a) Image acquisition protocol, including vertical stage displacement between the acquisition of time-lapse image, (b) trajectory reconstruction from a 20-second image sequence, and (c) quantities determined by dividing the visualisation field into stripes of width $\Delta y$. In (b), bacteria are first detected in each frame. Their successive positions between consecutive frames are then linked to reconstruct individual trajectories. In (c), we show one of the trajectory. Only trajectories longer than \qty{1}{\s} are retained in the analysis. First, the velocity of bacterium i at time $t$ is calculated as: $\mathbf{v}_i(t) = \delta \mathbf{x}_i/\delta t$ where $\delta \mathbf{x}_i$ is the displacement of the bacterium in the focal plane between two consecutive frames separated by a time interval $\delta t = \qty{100}{\ms}$. The velocities of bacteria whose positions have been localised within a stripe of width $\Delta y$ around a position y are then used to determine the net average velocity of the bacteria at that position and time $T$ (see Eq.~\ref{eq:vxcond}). The number of detected positions in each stripe is used to calculate $b(y,T)$. This quantity allows us to reconstruct the spatial bacterial density profile across the channel width and compute the diffusive velocity (see Eq.~\ref{eq:diffusive-velocity}).}
    \label{fig:proc}
\end{figure*}

\begin{figure*}[htpb]
    \centering
    \includegraphics[width = \linewidth]{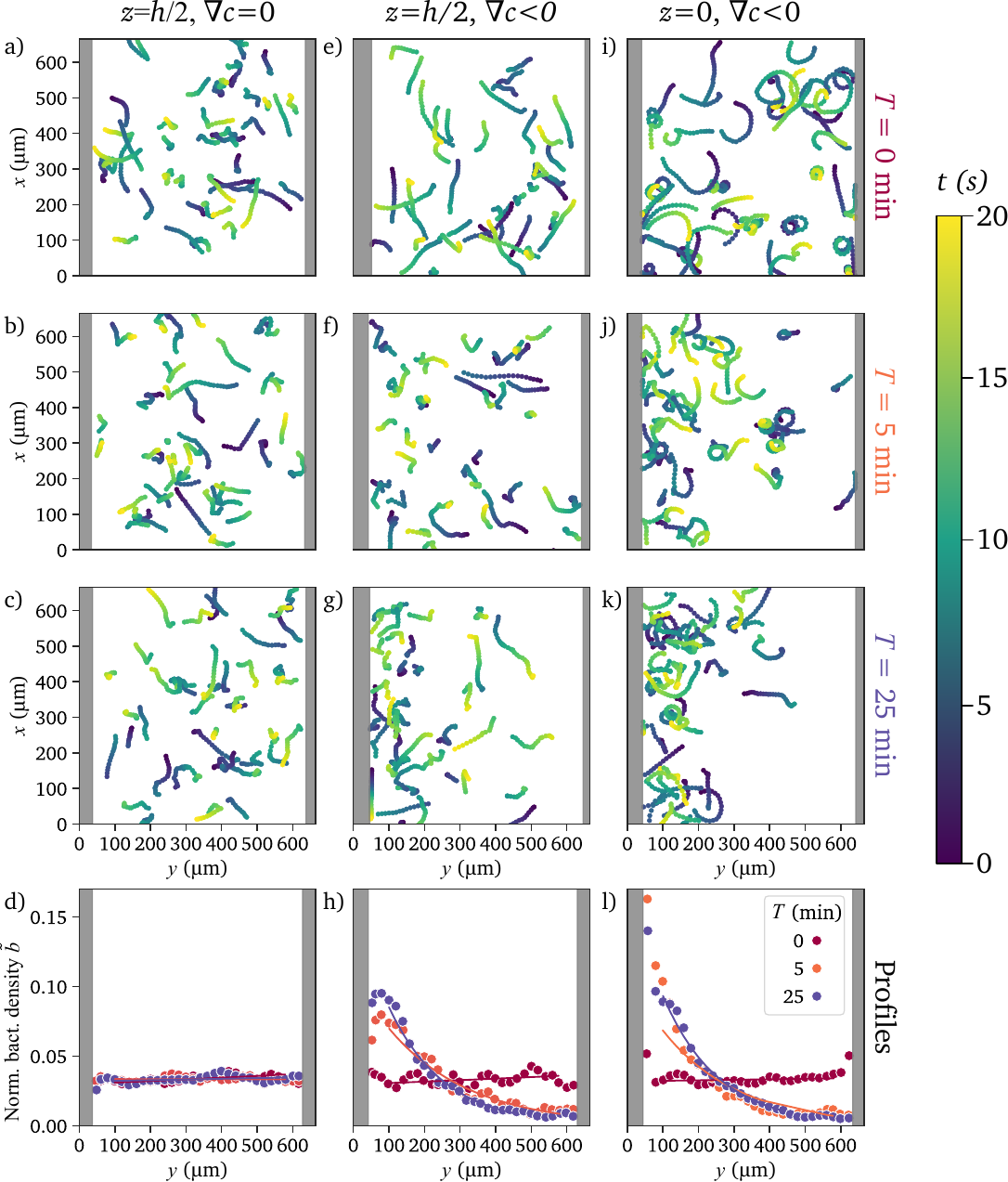}
    \caption{\textbf{Bacterial tracks and concentration profiles for two experiments.} Trajectories of \textit{E. coli} bacteria are shown at three different times $T$ during two experiments: without any gradient in the bulk $z=h/2$ (a-c), and with a $\nabla c =-\qty{200}{\uM\per\mm}$ MeAsp gradient either in the bulk $z=h/2$ (e-g) or on the bottom agar surface $z=0$ (i-k). The tracks are colour-coded with the time $t$ from \qtyrange{0}{20}{\s}. The corresponding bacterial profiles, measured at $T = \qty{0}{\minute}$ (\textcolor{spect0}{\ding{108}}), \qty{5}{\minute} (\textcolor{spect5}{\ding{108}}) and \qty{25}{\minute}(\textcolor{spect25}{\ding{108}}), are displayed at the bottom of each column (d,h,l).}
    \label{fig:tracks}
\end{figure*}

The different steps of the trajectories acquisition and analysis are schematised in Fig.~\ref{fig:proc}. The first step (see Fig.~\ref{fig:proc}a) involved recording time-lapse images (films) at seven intervals, \qtyrange{5}{15}{\minute} apart, at three different $z$ positions: $z = 0$ (on the agar), $z = h/2$ (half-height in the bulk) and $z = h$ (on the PDMS). The time at which a sequence of three films is acquired is noted $T$. The measurement of $T$ begins when the pump stops injecting bacteria into the central channel. Films were recorded at 10 frames per second at different times during the experiment. Most films lasted \qty{20}{\s} (i.e., \qty{200}{frames}), except for a few experiments that lasted \qty{60}{\s}. Bacterial tracks were computed from the sequences of phase fluorescent images using TrackMate~\cite{tinevez2017trackmate}, before being post-processed using an in-house Python code (see Fig.~\ref{fig:proc}b). Examples of tracks obtained are shown in Fig.~\ref{fig:tracks}. For each track $i$, the two-dimensional positions $\mathbf{x}_i(t) = (x_i(t),y_i(t))$ were measured, from which the two-dimensional velocities $\mathbf{v}_i(t) = \delta \mathbf{x}_i / \delta t$ using a sampling time $\delta t = \qty{0.1}{\s}$ were computed. Only tracks longer than one second, which correspond to trajectories made over ten ($x_i$,$y_i$) positions, were retained. Additionally, to separate motile bacteria from non-motile bacteria, the average velocity, $\bar{v}_i$ of each track was calculated and tracks with $\bar{v}_i < \qty{5}{\um\per\s}$ were removed.

The spatial distribution of bacteria $b(y,T)$ and the diffusion coefficient $\mu$ are essential to determine the chemotactic response (see Eq.~\ref{eq:vmu} and \ref{eq:sum}). To determine the profile $b(y,T)$, the data were spatially binned in the gradient direction over a stripe of width $\Delta y$, chosen as $\Delta y = \qty{40}{\um}$ because it is slightly longer than one correlation length of the bacterial trajectories ($l_c \sim v_s \tau \sim \qty{25}{\um}$). In order to fully neglect the effects of hydrodynamic coupling~\cite{lauga2009hydrodynamics} of swimming bacteria with surfaces, the profile is only measured \qty{100}{\um} away from surfaces. The diffusion coefficient $\mu$ is given by $\mu = v_s^2 \tau$~\cite{lovely1975statistical,taktikos2013motility,lauga2020}, where $\tau$ is the characteristic time of the exponential decrease of the velocity correlation functions, and $v_s$ is the swimming velocity defined as $v_s = \langle \lvert \mathbf{v}_i \cdot \mathbf{e}_y \lvert \rangle_i$, $\langle \cdot \rangle_i$ represents the average calculated across all tracks in a particular film~\cite{Gargasson2025}. In SM~\ref{sec:mot}, we present the swimming velocity and correlation time values obtained from our experiments, neither of which appears to be influenced by chemoattractant concentration. In practice, the values of the diffusion coefficient $\mu$ and swimming velocity $v_s$ have been recalculated for each film. This has the advantage of accounting for changes in motility between batches of bacteria. 

To determine the chemotactic velocity, we will assume (and subsequently verify in SM~\ref{sec:TimeSeparation}) that the flux along the $y$-axis is stationary over the duration of film acquisition, i.e., over \qty{20}{\s}. This assumption implies that the bacteria's spatial distribution profile changed little over the acquisition time. Under this assumption, we can use Eq.~\eqref{eq:ks} and rewrite it for the case of a 1D gradient along the $y$-direction as follows:
\begin{equation}
    v_c(y,T) = v(y,T) + \frac{\mu(T)}{b(y,T)} \frac{\partial b(y,T)}{\partial y}
\end{equation}
We defined the diffusive velocity as 
\begin{equation}
    v_\mu(y,T) = - \frac{\mu(T)}{b(y,T)} \frac{\partial b(y,T)}{\partial y}
    \label{eq:vmu}
\end{equation}
such that:
\begin{equation}
    v_c(y,T) = v(y,T) - v_\mu(y,T).
    \label{eq:sum}
\end{equation}
The diffusive velocity is determined by discretising the "local" gradient of $b(y,T)$. Specifically, the "local" gradient is calculated by finding the difference in bacterial population between the stripe below, $b(y - \Delta y, T)$, and the stripe above, $b(y + \Delta y, T)$ (see Fig.~\ref{fig:proc}c). This difference is then divided by the distance between the two stripes, which is $2 \Delta y$. The diffusive velocity is thus given by:
\begin{equation}
    v_\mu (y,T) = - \mu(T) \frac{b(y + \Delta y,T) - b(y - \Delta y,T)}{2 b(y,T) \Delta y}, 
    \label{eq:diffusive-velocity}
\end{equation}
and the net velocity by:
\begin{equation}
    v(y,T) = \left \langle \mathbf{v}_i(y_i) \cdot \mathbf{e}_y; \, y_i \in \left[ y - \frac{\Delta y}{2}, y + \frac{\Delta y}{2} \right] \right \rangle_{\Delta y}
    \label{eq:vxcond}
\end{equation}
In this equation, the brackets $\langle \cdot \rangle_{\Delta y}$ represent the average of all tracks whose positions fall within a stripe of width $\Delta y$ that is centred at $y$. The Fig.~\ref{fig:proc}c) schematises a specific trajectory and highlights the portion of the trajectory used to determine the contribution of this trajectory to net velocity. This operation is performed on all trajectories that intersect the stripe of width $\Delta y$ centered at $y$. The Eqs~\eqref{eq:diffusive-velocity}-\eqref{eq:vxcond} are applied to stripes where the number of bacteria $b(y)$ is more than one-tenth of the maximum of $b(y)$. The chemotactic velocity $v_c(y,T)$ is obtained by summing the net $v(y,T)$ and diffusive $v_\mu(y,T)$ velocities. The velocities can be averaged over $y$ to calculate the average chemotactic velocity $\bar{v}_c(T) = \langle v_c(y,T) \rangle_y$, the average diffusive velocity $\bar{v}_{\mu}(T) = \langle v_\mu(y,T) \rangle_y$, and the average net velocity $\bar{v}(T) = \langle v(y,T) \rangle_y$. The average velocities normalised by the swimming velocity $v_s(T)$ are noted $\tilde{v}_c(T)$, $\tilde{v}_\mu(T)$ and $\tilde{v}(T)$. Finally, $\tilde v_c (T)$ can be divided by $\nabla c$ (defined in Sec.~\ref{sec:chexp}) to obtain $\chi (T)$ for a specific film. 
In a similar manner, a "local" susceptibility $\chi(y,T)$ can be determined from the chemotactic velocity $v_c(y,T)$ measured at position y for the video acquired at time $T$ as:
\begin{equation}
\chi(y,T)=\frac{v_c(y,T)}{v_s(T)}\frac{1}{\nabla c}.
\label{eq:chi_local}
\end{equation} 
This quantity will be represented as function of the "local" concentration $c(y) = c_1 + (200 + y) \nabla c$ in Sec.~\ref{sec:transient_local}.
Note that the chemotactic susceptibility $\chi$ is calculated using the \textit{normalised} chemotactic velocity, meaning its unit is $[\chi] = \unit{\um\per\uM}$.
% This definition of $\chi$ allows for a more accurate comparison of bacterial batches with different $v_s$.

\subsection{Detection threshold of chemotactic velocity}
\label{sec:thresh_vel}

To assess the setup's reliability, four control experiments were conducted without a gradient, meaning that the concentrations in both lateral channels were equal ($c_1 = c_3$). Three of these experiments had a concentration of $c_1 = c_3 = 0$, while one experiment used a concentration of $c_1 = c_3 = \qty{42}{\mM}$ of casamino acids. The chemotactic velocity measured in these experiments was centred on \qty{0.1 \pm 0.7}{\um\per\s}, confirming the absence of chemotaxis or the existence of fluid flows along $y$. The \qty{95}{\percent} confidence interval of \qty{0.1 \pm 0.7}{\um\per\s} reflects the precision of our experimental setup. This \qty{0.7}{\um\per\s} corresponds to a chemotactic velocity lower than \qty{5}{\percent} of the swimming velocity $v_s$. Therefore, this threshold of $\qty{0.7}{\um\per\s} \approx \num{0.05}\,v_s$ is selected as the limit above which we can confidently assert that a significant chemotactic response occurs.

\section{Results}

The quantification of the chemotaxis of \textit{E. coli} bacteria towards MeAsp and casamino acids is carried out in a microfluidic chip consisting of three parallel channels, following the protocol of~\cite{Gargasson2025} inspired by~\cite{Ahmed_2010,Ahmed_Shimizu_Stocker_2010}, as depicted in Fig.~\ref{fig:setup}. Figure~\ref{fig:tracks} illustrates our observations after tracking the bacteria and reconstructing their trajectories from 20-second films. It presents two experiments conducted at $\nabla c =0$ and $-\qty{200}{\uM\per\mm}$ and shows results of the analysis of image sequences acquired at times, $T = \qtylist{0;5;25}{\minute}$, and for $z=h/2$ (Fig.~\ref{fig:tracks}e)-g) and $z=0$ (Fig.~\ref{fig:tracks}i-k).

When there is no chemical gradient, the bacterial trajectories show no significant evolution, as illustrated in Fig.~\ref{fig:tracks}a-c). The distribution of bacteria within the channel remains unchanged throughout the experiment; the population stays uniformly distributed at all times, as demonstrated in Fig.~\ref{fig:tracks}d). The observed profiles are consistent with the trajectories shown in Fig.~\ref{fig:tracks}a-c) which shows bacteria present throughout the channel with trajectories that move in all directions without any preferential direction.

When experiments are conducted using a uniform gradient of chemoattractant, the bacterial density profiles change over time. A significant accumulation of bacteria is observed on the side of the channel that is closest to the source with the highest concentration of chemoattractant. Trajectories of bacteria in a $\nabla c = \qty{-200}{\uM\per\mm}$ MeAsp gradient, recorded in the bulk $z = h/2$, are shown in Fig.~\ref{fig:tracks}e-g). This accumulation typically takes place over several minutes and reaches a stationary state characterised by an exponential concentration profile of bacteria, see Fig.~\ref{fig:tracks}h), as already reported in the literature~\cite{Kalinin_Jiang_Tu_Wu_2009,Ahmed_Shimizu_Stocker_2010,Ahmed_2010,Menolascina_Rusconi_Fernandez_Smriga_Aminzare_Sontag_Stocker_2017}. In those studies, the characteristic length $\lambda$ that is associated with the exponential decay of the profile $b(y)$ is used to estimate the chemotactic velocity, $v_c$, using the relation $v_c = \mu / \lambda$. This method requires establishing a stationary profile, which may take several minutes to achieve. In Sec.~\ref{sec:chemo_vel}, we will demonstrate that the chemotactic velocity can be measured from any image sequence acquired at any time by analysing the net and diffusive velocities computed from the bacteria trajectories.
% Each experiment typically generates data for one chemotactic pairing of $\bar{c}$ and $\nabla c$.
%\com{These profiles are analysed to determine and quantify chemotaxis, either by adjusting the profile exponentially~\cite{Gargasson2025} or by estimating the chemotactic migration coefficient (CMC)~\cite{Kalinin_Jiang_Tu_Wu_2009}, which characterises its asymmetry.}
%We will then assess the accuracy of this method in measuring the chemotactic velocity. 
Following that, we will show in Sec.~\ref{sec:transient_local} how the method can be applied to determine a "local" chemotactic velocity. Furthermore, all our measurements will be used to construct the chemotactic susceptibility curve $\chi = f(c)$. Finally, we will study bacterial chemotaxis on surfaces in Sec.~\ref{Sec:surface} by analysing image sequences recorded at different $z$ positions in the channel, $z = 0$, $z = h/2$, and $z = h$.

\subsection{Chemotactic velocity}
\label{sec:chemo_vel}

The method we propose for quantifying the chemotactic velocity of bacteria is based on calculating "local" bacterial fluxes derived from analysing their trajectories. The net flux is decomposed into a chemotactic and a diffusive part. Conservation of fluxes allows us to obtain the chemotactic velocity as expressed by Eq.~\eqref{eq:sum}, which can be spatially averaged over $y$, yielding: $\bar{v}_c(T) = \bar{v}(T) - \bar{v}_\mu(T)$. After normalisation by the swimming velocity, we obtain $\tilde{v}_c(T) = \tilde{v}(T) - \tilde{v}_\mu(T)$.

\begin{figure}[hbtp]
    \centering
    \includegraphics[width = \linewidth]{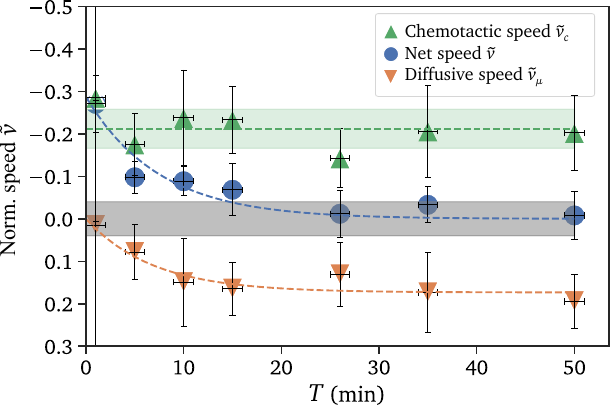}
    \caption{\textbf{Evolution of the average net, diffusive and chemotactic velocities in a typical experiment measured at $z = h/2$}. All velocities are normalised by the swimming velocity $v_s(T)$: blue circles correspond to the net velocity $\tilde{v}$, orange downward pointing triangles to the diffusive velocity $\tilde{v}_\mu$, and green upward pointing triangles to the chemotactic velocity $\tilde{v}_c$. The vertical bars indicate the standard deviation of the values on the stripes, while the horizontal bars correspond to film acquisition duration. The net and diffusive velocities are fitted by an exponential function $\tilde{v} \propto \exp(-t / \tau^\mathrm{stat})$ (dashed lines), with $\tau^\mathrm{stat}$ the characteristic time to reach the stationary state: $\tau^\mathrm{stat} \simeq \qty{7}{\minute}$ for the net velocity, and $\tau^\mathrm{stat}_\mu \simeq \qty{9}{\minute}$ for the diffusive velocity. Green strip: average and standard deviation of each average chemotactic velocity value. Grey strip: threshold of $\tilde{v}_c$ over which chemotaxis is detected. This experiment was performed using MeAsp with $\bar{c} = \qty{100}{\uM}$ and $\nabla c = \qty{200}{\uM\per\mm}$.}
    \label{fig:velocities}
\end{figure}

In Fig.~\ref{fig:velocities}, we have represented the net velocity $\tilde{v}(T)$ and the diffusive velocity $\tilde{v}_\mu(T)$ measured at seven different times $T$ for the experiment depicted in Fig.~\ref{fig:tracks}. The net velocity is at its highest absolute value at the beginning of an experiment, see blue circles in Fig.~\ref{fig:velocities}.
%for the evolution of the normalised net velocity $\tilde{v}=v/v_s$. 

This high negative value indicates that the bacterial population is moving towards the source of the attractant. As time goes on, the absolute net velocity decreases and approaches a value close to zero after \qty{25}{\minute}. This decrease is concomitant with an increase in the normalised diffusive velocity $\tilde{v}_\mu$ (orange downward pointing triangles), calculated from the concentration profiles of bacteria as they become more and more uneven. After \qty{25}{\minute}, this velocity stabilises to a plateau value as the profile reaches the exponential stationary state (see Fig.~\ref{fig:tracks}h). This stationary state is achieved when the diffusive flux is important enough to compensate the chemotactic flux, resulting in a total flux and a net velocity $\tilde{v}$ of zero. As highlighted in Eq.~\eqref{eq:sum}, the sum of the net and diffusive velocities at all times $T$ gives the chemotactic velocity $\tilde{v}_c$, see green upward pointing triangles in Fig.~\ref{fig:velocities}. The chemotactic velocity is almost constant over time, which is consistent with the existence of a steady uniform gradient along the width of the central channel where the bacteria are located, with a mean value close to \num{-0.2} well above the detection threshold (grey strip), validating our method.

\subsection{Transient and local measurements improve measurement of chemotactic susceptibility}
\label{sec:transient_local}

%The proposed method, as demonstrated in the previous section, offers an efficient approach to measuring the chemotactic velocity at any time during the experiment. 
In this section, we will demonstrate that the chemotactic velocity and, consequently, the chemotactic susceptibility can be measured in each stripe and at each time $T$. We will demonstrate how this refined methodology enables more accurate quantification of the chemotactic response, leading to better statistical analysis and greater reliability of the findings.
%By locally measuring the net velocity and calculating the diffusive velocity from the bacterial concentration profile, we can determine the chemotactic velocity and, consequently, the chemotactic susceptibility at any given time. This method contrasts with the stationary approach, which only allows for the determination of a single chemotactic velocity from one experiment. 
%First, we will first focus on the temporal aspect and its dependence on the variable $y$, while the impact of the height $z$ will be addressed in the following section.
First, we will emphasise a key advantage of our method: measurements can be made from the very first recorded time-lapse, eliminating the need to wait for a stable bacterial concentration profile to develop. 

\begin{figure*}[htpb]
    \centering
    \includegraphics[width = \linewidth]{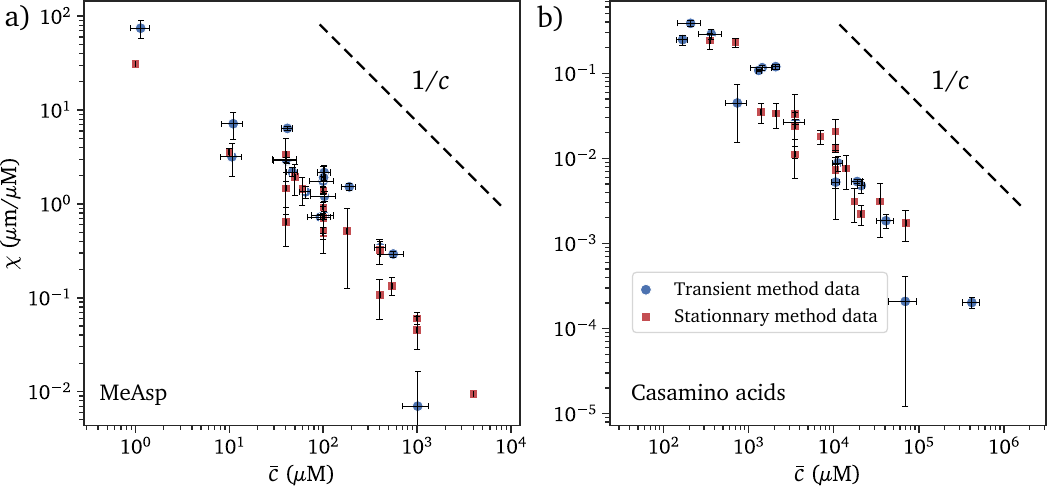}
    \caption{\textbf{Measuring the bacterial velocity bias in the transient state allows for a much quicker quantification of the chemotactic bias.} Chemotactic susceptibility $\chi(\bar c)$ measured with (a) MeAsp and (b) casamino acids as a function of the average concentration in the channel $\bar c$. Blue circles: susceptibility deduced in the transient state ($T \leq \qty{3}{\minute}$) from the bias in bacterial velocities along the gradient. Red squares: susceptibility deduced from the steady spatial profile of the bacteria~\cite{Gargasson2025}.}
    \label{fig:chi(c)_transient}
\end{figure*}

To demonstrate this advantage, the first time-lapse images of each experiment recorded halfway between the surfaces ($z = h/2$) were analysed with the method described in Sec.~\ref{sec:analyse}. In the early stage of the transient state, the bacterial concentration in the channel remains roughly uniform across the channel width (see Fig.~\ref{fig:tracks}h). The diffusive velocity accounts for less than \qty{1}{\percent} of the swimming velocity, leading to $\tilde{v}_c \approx \tilde{v}$ and $\chi(T)\approx \tilde{v}/\nabla c$.
The values of $\chi(T \leq \qty{3}{\minute})$ are plotted as a function of the average chemical concentration in the channel $\bar{c}$, see blue circles in Fig.~\ref{fig:chi(c)_transient}. The data points from the stationary profile analysis are shown as red squares. For both methods, we obtain a clear $1/\bar{c}$ evolution of the chemotactic susceptibility $\chi(\bar{c})$, characteristic of a log-sensing response, over more than three decades. 

\begin{figure*}[htpb]
    \centering
    \includegraphics[width = \linewidth]{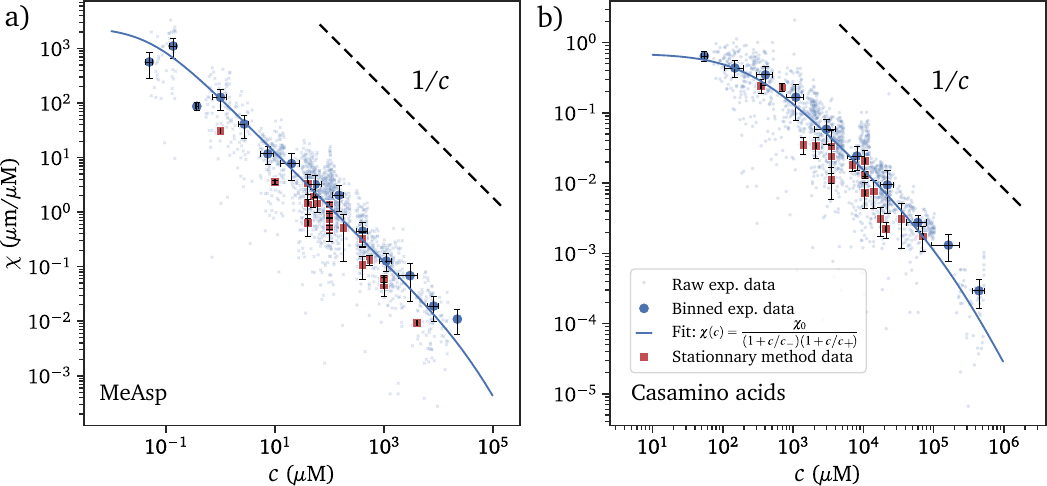}
    \caption{\textbf{Transient and local measurements allow for a quicker and more accurate quantification of the log-sensing response.} Chemotactic susceptibility of chemoattractants (a) MeAsp and (b) casamino acids, respectively, as a function of their "local" concentration $c$. Each small shaded blue circle corresponds to the value of the chemotactic susceptibility $\chi(Y,T)$ measured for each stripe $Y$ of "local" concentration $c(Y)$ at each time $T$ for each experiment, while the larger blue circles with error bars correspond to their binning in $\log(c)$. Data from the steady spatial profile of the bacteria~\cite{Gargasson2025} are shown as \textcolor{cred}{\ding{110}}, one point corresponding to only one experiment. Solid lines: fit of the data as $\chi(c) = \chi_0/((1+c/c_-)(1+c/c_+))$, with (a) $\chi_0 = \qty{2.5(20)e3}{\um\per\uM}$, $c_- = \qty{5(3)e-2}{\uM}$ and $c_+ = \qty{5(3)e4}{\uM}$ for MeAsp, and (b) $\chi_0 = \qty{0.7(4)}{\um\per\uM}$, $c_- = \qty{2(1)e2}{\uM}$ and $c_+ = \qty{2(1)e5}{\uM}$ for casamino acids. Dashed lines: $\chi(c) \propto 1/c$.}
    \label{fig:chi(c)}
\end{figure*}

The second strength of the proposed method is its ability to obtain a "local" measurement of the chemotactic velocity by analysing a specific portion of the entire image. The channel containing the bacteria is divided into stripes of width $\Delta y$ for the analysis. In each stripe, we evaluate the chemotactic velocity $v_c(y,T)$ by analysing the detected trajectories within that stripe (see Sec.~\ref{sec:analyse} and Fig.~\ref{fig:proc}c). Using this analysis, we can calculate the "local" chemotactic susceptibility as $\chi(y,T)$, by applying Eq.~(\ref{eq:chi_local}). Fig.~\ref{fig:chi(c)} presents the values of $\chi(y,T)$ as a function of the "local" concentrations of chemoattractants: MeAsp (a) and casamino acids (b). Each light blue circle represents a data point obtained from the \num{30} stripes across \num{7} different time points, $T$, for all $n$ experiments, where $n = \num{24}$ for MeAsp and $n = \num{21}$ for casamino acids. A logarithmic binning in $c$ was applied to obtain the dark blue circles. As in Fig.~\ref{fig:chi(c)_transient}, a clear $1/c$ trend appears over multiple decades in concentration for both chemoattractants. This log-sensing response approximately spans from \qtyrange{1e0}{1e4}{\uM} for MeAsp, and from \qtyrange{5e2}{5e5}{\uM} for casamino acids. Outside of these bounds, the chemotactic response seems to deviate from the $1/c$ trend. We, thus, fit our data using Eq.~\eqref{eq:SiTu}, $\chi(c) = \chi_0/\left((1 + c/c_-)(1 + c/c_+)\right)$, which includes a lower limit ($c_-$) and an upper limit ($c_+$) to the log-sensing regime~\cite{Si_Wu_Ouyang_Tu_2012,Menolascina_Rusconi_Fernandez_Smriga_Aminzare_Sontag_Stocker_2017,Cremer2019,alert2022cellular}. For both chemoattractants, the best fit of the data yields the values of both concentration thresholds $c_-$ and $c_+$ as well as the coefficient $\chi_0$. In practice, those thresholds are found close to the lowest and highest concentration values beyond which the chemotactic velocity becomes lower than the detection threshold determined in Sec.~\ref{sec:analyse}. For MeAsp, we find $\chi_0 = \qty{2.5e3}{\um\per\uM}$, $c_- = \qty{5e-2}{\uM}$ and $c_+ = \qty{5e4}{\uM}$ (see Fig.~\ref{fig:chi(c)}a), while for casamino acids, we find $\chi_0 = \qty{0.7}{\um\per\uM}$, $c_- = \qty{2e2}{\uM}$ and $c_+ = \qty{2e5}{\uM}$ (see Fig.~\ref{fig:chi(c)}b). The results for MeAsp are consistent with the study by Kalinin \textit{et al.} who found a log-sensing chemotactic response between \qtyrange{5e0}{1e4}{\uM}~\cite{Kalinin_Jiang_Tu_Wu_2009}. The chemotactic behaviour in response to a casamino acids gradient also show a similar log-sensing response.

Our data can also be compared with the data obtained with the stationary method~\cite{Kalinin_Jiang_Tu_Wu_2009,Menolascina_Rusconi_Fernandez_Smriga_Aminzare_Sontag_Stocker_2017,Gargasson2025} (red squares in Fig.~\ref{fig:chi(c)}).
%In order to compare our method with the stationary method used in the literature~\cite{Kalinin_Jiang_Tu_Wu_2009,Menolascina_Rusconi_Fernandez_Smriga_Aminzare_Sontag_Stocker_2017,Gargasson2025}, the chemotactic susceptibility derived from the stationary bacterial profile of our experiments is presented as red squares in Fig.~\ref{fig:chi(c)}.
The first striking observation is the sheer difference in the quantity of data generated by each method: for the $n = \num{23}$ experiments conducted with a gradient of MeAsp, more than \num{2e5} data points are retrieved with our method compared to only $\approx \num{80}$ with the stationary method. Our method also significantly broadens the range of concentrations at which a chemotactic velocity is observed. For MeAsp, the stationary method detects a chemotactic response for concentrations ranging from \qty{1e0}{\uM} to \qty{6e3}{\uM}. In contrast, our method allows observation of chemotaxis over a wider concentration range, from \qtyrange{5.0e-2}{1.5e4}{\uM}.

\subsection{Chemotactic drift on surfaces}
\label{Sec:surface}

After quantifying the chemotactic behaviour of \textit{E. coli} in the bulk, we will now focus on its chemotactic response near surfaces, specifically shifting from the analyses in Sec.~\ref{sec:chemo_vel} and \ref{sec:transient_local}, which were based on data recorded at $z = h/2$, to films taken on the top and bottom surfaces. As the focal plane depth is roughly \qty{40}{\um}, bacteria located within \qty{20}{\um} of the surface are detected~\cite{Gargasson2025}.

The behaviour observed on surfaces is consistent with that at $z = h/2$: The bacterial population profile across the channel (see Fig.~\ref{fig:tracks}l) transitions from a flat distribution, indicating a homogeneous distribution of bacteria, to one that shows an accumulation of bacteria near the source of the chemoattractant. This pattern mirrors what is observed at $z = h/2$, as shown in Fig.~\ref{fig:tracks}h). When examining individual trajectories, we find that bacteria on the surfaces exhibit more circular movement (see Fig.~\ref{fig:tracks}i-k) compared to the straighter trajectories observed at $z = h/2$ (see Fig.~\ref{fig:tracks}e-g). The presence of circular trajectories on surfaces has been previously documented and can be explained by the hydrodynamic interactions between bacteria and surfaces. These interactions come from viscous forces exerted by the displaced fluid on the swimming bacteria, which are induced by the rotation of the flagella and the bacterial body~\cite{lauga2009hydrodynamics}.

\begin{figure*}[htpb]
    \centering
    \includegraphics[width = \linewidth]{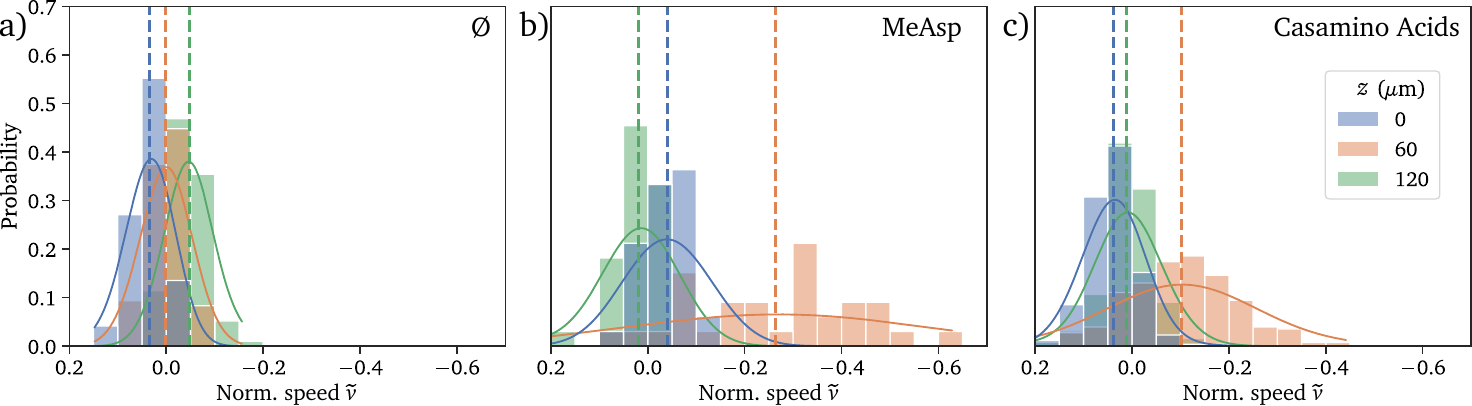}
    \caption{\textbf{Absence of chemotactic drift on surfaces.} Distribution of the normalised net velocity $\tilde{v}(y,T) = v(y,T) / v_s(T)$ for $T \leq \qty{3}{\minute}$, with (a) no chemoattractant gradient, (b) a MeAsp gradient of \qty{-0.2}{\uM\per\um}, and (c) a casamino acids gradient of \qty{-0.7}{\uM\per\um}. Blue and green histograms show bacteria velocities measured on surfaces, respectively the agar ($z=0$) and PDMS ($z=h$) surfaces. Orange histograms show the velocity distribution in the bulk at $z = h/2$, halfway between the surfaces. A clear chemotactic drift appears in the bulk when a gradient of chemoattractants is established, while no drift is noticeable on the surfaces. 
    Solid lines: Kernel density plots.}
    \label{fig:velocities_surface}
\end{figure*}

The chemotactic velocities $v_c(y,z=0,T \leq \qty{3}{\minute})$ and $v_c(y,z=h,T \leq \qty{3}{\minute})$ were then evaluated. For $T \leq \qty{3}{\minute}$, the chemotactic flux is roughly equal to the net flux as the concentration profile of bacteria is nearly uniform. The distributions of the normalised net velocities $\tilde{v}(y)$ are thus shown in Fig.~\ref{fig:velocities_surface} for the three different $z=0$, $h/2$ and $h$. In the experiments conducted without chemical gradients, as illustrated in Fig.~\ref{fig:velocities_surface}a), the distributions measured on the surfaces (represented by the blue and green bins) and at half-height in the cell (represented by the orange bins) overlap. These distributions can be fitted with a Gaussian distribution that has a mean of zero and a root mean square deviation of $\approx \num{0.05}$. In the presence of a chemoattractant, and for the measurements taken at $z=h/2$ (shown in orange bins in Fig.~\ref{fig:velocities_surface}b) and c), the distributions are broader compared to those measured without the attractant and with a negative mean of respectively \num{-0.25} and \num{-0.1}. This indicates the chemotactic movement of bacteria towards the source of the chemoattractant, as discussed in the previous sections. The shift and enlargement of the distributions are not observed in the measurements taken on the surfaces (green and blue bins in Fig.~\ref{fig:velocities_surface}b) and c). The distribution retains the same shape as that measured without a chemoattractant (see Fig.~\ref{fig:velocities_surface}a), showing a zero average and a root mean square deviation of $\approx \num{0.05}$, clearly suggesting that the surfaces inhibit chemotaxis. 

This fascinating effect of surfaces on chemotactic behaviour is consistent with a previous study on \textit{Caulobacter crescentus}' chemotaxis towards MeAsp, which showed a zero net velocity close to surfaces~\cite{Grognot_2021}.

\section{Discussion and Conclusions}

Our research presents a novel technique for determining the chemotactic response of bacteria to a chemical gradient. Our approach involves analysing bacterial tracks extracted from time-lapse images to determine the chemotactic velocity. We assess the net and diffusive velocities to characterise bacterial movement in response to a chemical gradient. Accurately evaluating the diffusive flux $\mu(c)\partial b/\partial y$ can be quite challenging~\cite{Bouvard_2022}, as bacterial motility may depend on the "local" concentration $c$ or the concentration experienced over a specific duration. However, the weak dependence of the bacterial diffusivity $\mu$ on the concentration of chemoattractants used in this study (see SM~\ref{sec:mot}) helps to perform the evaluation consistently. We thus successfully apply our method to experiments performed on a 3-channel microfluidic chip, where a population of bacteria can be placed in a steady, uniform gradient of chemical elements.

By using this method, we assess the chemotactic susceptibility of \textit{E. coli} to MeAsp and casamino acids. The data collected using MeAsp can be compared to existing research~\cite{Ahmed_2010, Kalinin_Jiang_Tu_Wu_2009}. Additionally, the use of casamino acids broadens the approach to include mixed consumable small peptides, which mimic the natural nutrient gradients bacteria encounter in their environments. Our measurements are compared with those obtained from analysing the steady-state bacteria concentration profile. We demonstrate that the proposed method is (i) faster, as it does not require waiting for the bacterial concentration profile to reach a stationary state, which takes around twenty additional minutes after the establishment of the chemical gradient ($\simeq \qty{30}{\minute}$), see SM~\ref{sec:steady}; (ii) allows for measurements within a smaller spatial window, facilitating localised analysis of chemotaxis; and (iii) offers improved statistics because more data can be collected during a single experiment. Thanks to these advantages, the concentration range in which a chemotactic response is observed is identified with precision. 

For both chemoattractants, the data can be adjusted using Eq.~\eqref{eq:SiTu}, which simplifies to $\chi(c)=\chi_0/c$ when $c_- \ll c \ll c_+$, an evolution consistent with the theoretical analyses of Keller \& Segel~\cite{Segel_1977}. According to this scaling law, the chemotactic velocity varies as $\log(c)$, indicating a log-sensing response. This confirms the results previously obtained with MeAsp in a similar microfluidic setup by Kalinin \textit{et al}~\cite{Kalinin_Jiang_Tu_Wu_2009} and invalidates the scaling law used by Ahmed \textit{et al}~\cite{Ahmed_Shimizu_Stocker_2010}. In the latter case, the difference can be attributed to the narrower concentration range of \qtyrange{0.1}{1}{\mM} used compared to our study, which ranges from \qtyrange{0.0001}{100}{\mM}. This larger range allows for a more accurate fit of the scaling laws of $\chi(c)$, as well as the determination of the bounds to the log-sensing regime, which could help shed some light on the biochemical processes involved in the detection of chemicals. 

Our method was applied to image sequences acquired on surfaces, showing that there is no chemotaxis on the surfaces. We believe this effect can be explained by the shapes of the trajectories bacteria take near surfaces. Due to hydrodynamic interactions between the bacteria and the surface, their trajectories become circular~\cite{lauga2009hydrodynamics}. This circular motion leads to a reorientation of the swimming direction of the bacteria. Importantly, this reorientation occurs more quickly than the time needed for a chemical signal to provoke a change in the swimming direction of the bacteria in response to variations in chemoattractant concentration. For \textit{E. coli}, tumbling allows the bacteria's body to reorient itself, allowing the bacteria to move away from the surface~\cite{Junot_2022}. As hydrodynamic interaction force decreases with distance from the surface, bacteria that move away from the surface experience a reorientation time defined by the run-and-tumble swimming mechanism. This allows for chemotaxis to occur again, enabling the bacteria to migrate toward the source of the chemoattractant. The reorientation mechanism of the bacteria enables them to return to the surface, positioning themselves closer to the source of the chemoattractant this time. This sequence of trapping, escaping, and chemotactic drift results in the distribution of bacteria on the surfaces evolving towards an exponential profile similar to that measured at half-height far from surfaces. This result confirms the inhibitory role of surfaces observed by Grognot and Taute~\cite{Grognot_2021} when analysing 3D trajectories of \textit{E. coli} bacteria in a concentration gradient.

\section{Perspectives}

One advantage of the method described is that it can be applied to bacteria that respond to chemical gradients differently than \textit{E. coli}. 
%For instance, some marine bacteria, such as \textit{V. alginolyticus}, exhibit an increase in swimming velocity with higher sodium concentrations \jb{Ref?}. 
%can also benefit from this method. 
%On the other hand, the statistical analysis confirmed that there is no correlation between the swimming velocity $v_s$ and diffusion coefficient $\mu$ with either $c$ or $\nabla c$ for \textit{E. coli}. Thus, the absence of chemokinetic behaviour is evident. 
In our method, the velocity $v_s$ and the diffusion coefficient $\mu$ are calculated from each image sequence, enabling us to account for these effects. Our approach also allows us to distinguish between different populations of bacteria within the suspension. For instance, we can filter out a subpopulation of non-motile bacteria from the motile ones.

The statistical analysis of bacterial trajectories significantly reduces the experimental time needed to determine chemotactic velocity. We believe that our proposed method will aid in the development of new, faster, and more precise techniques for screening bacterial molecule couples. Finally, the absence of chemotaxis on surfaces opens up interesting perspectives concerning chemotaxis in porous media.

\section*{Author Contributions}

P.M., H.A. and C.D. conceptualised the research. A.G., J.B., H.A. designed the experiments. J.B. conducted early experiments. A.G. performed the experiments and analysed the data. P.M. and H.A. secured funding. P.M., C.D. and H.A. supervised the project. A.G., J.B. and H.A. wrote the original draft of the manuscript. All authors reviewed the manuscript.

\section*{Conflicts of interest}

The authors have no conflict to declare.

\section*{Acknowledgements}

We thank Akash Ganesh and Gaëlle Lextrait for experimental help, as well as Frédéric Moisy for fruitful discussions. 
This work is supported by the French National Research Agency (ANR) through the “Laboratoire d’Excellence Physics Atom Light Mater” (LabEx PALM) as part of the “Investissements d’Avenir” program (ANR-10-LABX-0039), and by the CNRS through the Mission for Transversal and Interdisciplinary Initiatives (MITI), part of the 80 Prime program (RootBac project).

%%%END OF MAIN TEXT%%%


%\balance


%%%REFERENCES%%%
\bibliographystyle{rsc} %the RSC's .bst file
\bibliography{reference} %You need to replace "rsc" on this line with the name of your .bib file

\clearpage
\appendix

\section*{Supplementary Material}

\section{Characterisation of the concentration gradient between the channels}
\label{sec:steady}

\begin{figure}[htpb]
    \centering
    \includegraphics[width = 0.97 \linewidth]{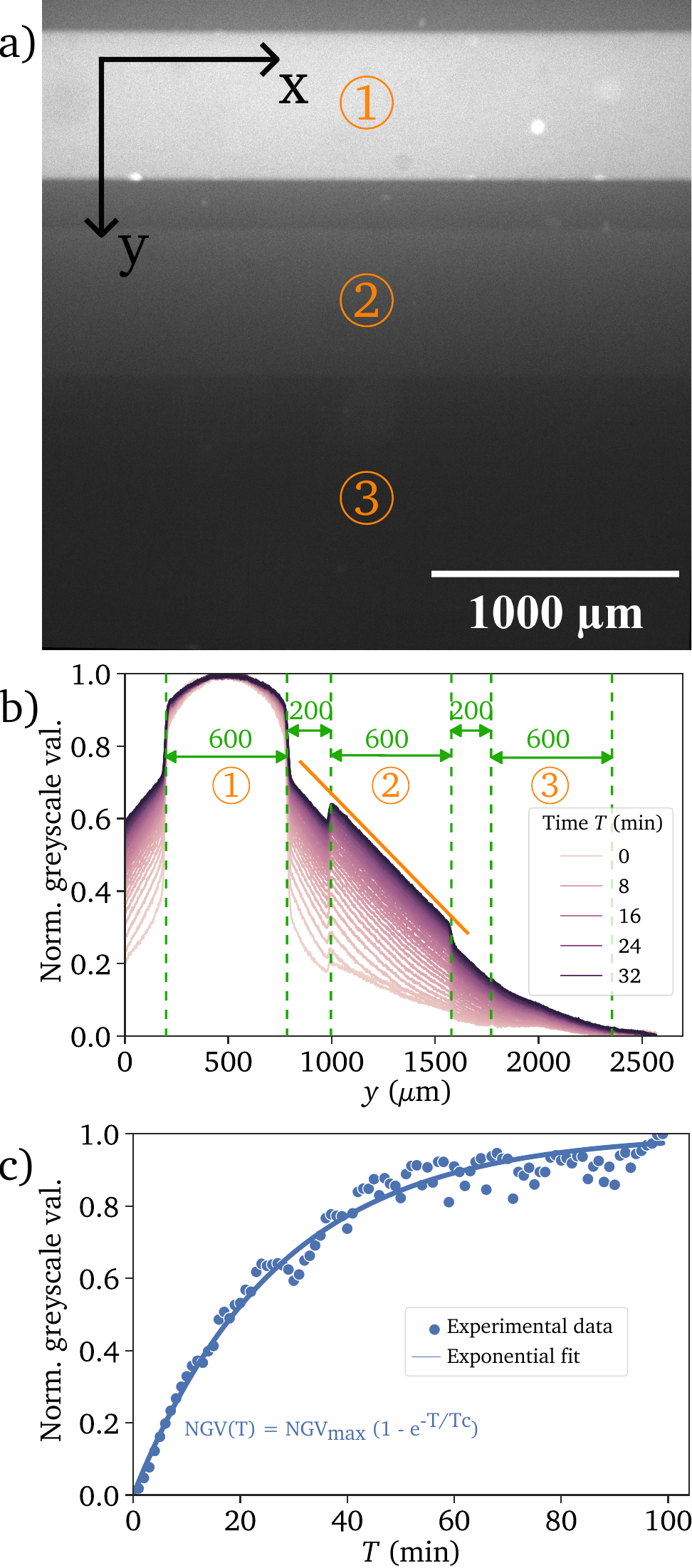}
    \caption{a) Image showing the three channels after $T = \qty{100}{\minute}$. The image was captured under fluorescence at $5 \times$ magnification. b) Spatial profile of the normalised greyscale value averaged over $x$ measured at different times. c) Time evolution of the normalised greyscale value in channel \chantwo. Fluorescein solution is injected into channel \chanone~at $T = 0$ at a rate of \qty{1}{\uL\per\minute}. Pure water is injected into channel \chanthree~at $T = 0$ at the same rate of \qty{1}{\uL\per\minute}. The flow is stopped in channel \chantwo~at $T = 0$. The time evolution is fitted with an exponential law: $\mathrm{NGV}(T) = \mathrm{NGV}_\mathrm{max}(1-e^{-T/T_c})$, with $T_c = \qty{24}{\minute}$ the characteristic diffusion time.}
    \label{fig:fluo}
\end{figure}

To characterise the gradient between channel \chanone~and channel \chanthree, we repeated the experimental protocol, replacing the chemoattractant with fluorescein. Additionally, a lower magnification objective was used to enable simultaneous visualisation of all three channels. Fig.~\ref{fig:fluo}a) shows an image acquired after several minutes of injecting dye into channel \chanone~and dye-free water into channel \chanthree. 

To quantify the dye concentration profile across the two channels, the image greyscale intensity was averaged along the $x$-direction. Figure \ref{fig:fluo}b) illustrates the profiles obtained by analysing an image sequence that begins when the flow in channel \chantwo~is stopped, with a flow rate of \qty{1}{\uL\per\minute} maintained in channels \chanone~and \chanthree. We observe that the light intensity in channel \chantwo~increases over time. Additionally, a gradient forms and, after a few minutes, stabilises into a constant gradient characterised by a linear evolution in light intensity across channel \chantwo. Meanwhile, the light intensity in channel \chanone~remains unchanged, whereas that in channel \chanthree~shows a slight increase. To characterise the time required for the intensity profile to establish, we calculated the average light intensity in channel \chantwo. This value is plotted in Fig.~\ref{fig:fluo}c) as a function of time.

We will now compare our observations with the analytical solution for the concentration profile between two boundaries separated by a distance $L$, where the concentrations are fixed at $c_1$ (at $y = 0$) and $c_3$ (at $y = L$).

The variation in concentration between channel \chanone~and channel \chanthree~follows Fick's law:
\begin{equation}
    \partial_t c = -D \partial^2_y c
    \label{eq:consc}
\end{equation}
The solution of this equation for a finite domain [0,L] with Dirichlet boundary conditions:
\begin{align}
    c(0,t) &= c_1;\\
    c(L,t) &= c_3
    \label{eq:dirich}
\end{align}
and an initial condition:
\begin{equation}
    c(x,0) = 0
\end{equation}
can be derived using a Fourier series expansion~\cite{crank1979mathematics}. At short times: The first term of the Fourier series dominates the dynamics, and the solution can be approximated by:

\begin{equation}
    \frac{c(y,t)-c_1}{c_3-c_1} \approx \frac{y}{L} + \frac{2}{\pi}\sin\left(\frac{\pi y}{L}\right) \cdot e^{-D^\mathrm{fluo}_\mathrm{med}\left(\frac{\pi}{L}\right)^2 t},
\end{equation}
where $D^\mathrm{fluo}_\mathrm{med}$ is the diffusion coefficient of fluorescein in the medium. The average concentration between the two channels: $\bar{c}(t) = (1/L)\int_0^L c(y,t)dy$ gives: 

\begin{equation}
    \frac{\bar{c}(t) - c_1}{c_3 - c_1} \approx \frac{1}{2} + \frac{4}{\pi^2} e^{-D^\mathrm{fluo}_\mathrm{med}\left(\frac{\pi}{L}\right)^2 t}
\end{equation}
The characteristic time required to reach a steady-state regime, where the average concentration $\bar{c}$ remains constant, is $L^2/(\pi^2 D)$. This time can be determined experimentally by analysing the curve that illustrates the changes in average light intensity in the central channel, as shown in Fig.~\ref{fig:fluo}c). The exponential fit (solid line in Fig.~\ref{fig:fluo}c) indicates a time of \qty{24}{\minute}. For a length $L = \qty{1000}{\um}$, this results in a molecular diffusion coefficient $D^\mathrm{fluo}_\mathrm{agar} = \qty{70}{\square\um\per\s}$, which is lower than the value of $D^\mathrm{fluo}_\mathrm{water} = \qty{430}{\square\um\per\s}$ reported for diffusion in water~\cite{Mustafa1993,Richbourg2021}. 

Several factors explain the reduction of the diffusion coefficient of the fluorescein observed in agar. These include interactions between the hydrogel network and the solute, as well as obstacles to solute transport created by the hydrogel matrix~\cite{Amsden1998,Masaro1999,Axpe2019}.

Note that the concentration gradient $\nabla c$ used throughout the article was calculated based on the concentrations $c_1$ and $c_3$ in the lateral channels where a constant flow is maintained, i.e. $\nabla c = (c_3 - c_1) / L$. Thus, $\nabla c$ is based on the assumption that a perfectly linear concentration profile is established between channels \chanone~and \chanthree. The concentration profile of fluorescein in Fig.~\ref{fig:fluo}b) suggests that this formula overestimates the gradient by \qty{40}{\percent}. However, this concentration profile, whose main objective is to confirm the linearity of the steady gradient, has itself a few caveats, e.g. the integration of light intensity through the $z$-axis, which is why we did not correct the theoretical value of the gradient when calculating the chemotactic susceptibility $\chi$.

The chemoattractants used in our experiments are not visible under fluorescence. We will assume that molecular size is the most critical factor determining the diffusion coefficient. If the molecules are of comparable size, we will further assume that they diffuse through the hydrogel with the same dynamics. To estimate the size of the molecules in the casamino acids mixture, we considered it to be composed of an equimolar distribution of \num{20} amino acid molecules. Based on this assumption, we calculated an average Stokes radius of approximately \qty{4.2}{\Angst}. MeAsp is an amino acid. Its radius can be estimated from aspartate's radius as $\approx \qty{4}{\Angst}$~\cite{Liu2020}. These values are comparable to the size of the fluorescein molecule ($\approx \qty{5}{\Angst}$)~\cite{Mustafa1993,Richbourg2021}. 

It is important to note that chemoattractant consumption by bacteria is neglected in Eq.~\eqref{eq:consc}. This is obvious for MeAsp, which is not metabolised by \textit{E. coli}. However, casamino acids are. The approximation will thus be only valid if the consumption rate of the constituents of the casamino acids by the bacteria does not exceed the rate at which they are replenished by diffusion from the lateral channels. 

The casamino acids are consumed at a rate below \qty{0.05}{\uM\per\OD\per\s}~\cite{ganesh:tel-03999273}. In our experiments, $\mathrm{OD} \approx \num{0.08}$, and $\bar{c} < \qty{3e5}{\uM}$ yielding a characteristic consumption time of \qty{1.25e6}{\min}. This duration is significantly longer than the $\sim \qty{24}{\minute}$ required to establish a gradient. Therefore, the metabolisation of the casamino acids can be neglected as well. 

Another critical timescale is the time required for chemoattractants to diffuse from the agar into channel \chantwo~and for a gradient to establish across the width of the channel. The transition from a state in which the channel contains no chemoattractant to one in which a gradient is fully established occurs via molecular diffusion. The characteristic time for this process is $\tau_D = h^2/(2 D^\mathrm{chemo}_\mathrm{water}) \approx \qty{15}{\s}$, with $D^\mathrm{chemo}_\mathrm{water} \approx \qty{500}{\square\um\per\s}$ the diffusivity of the chemoattractants in water based on their Stokes radii. This timescale is \num{100} times shorter than the time required to establish the gradient within the agar layer between channels \chanone~and \chanthree~(see Fig.~\ref{fig:setup}). This difference arises from the aspect ratio: the channel height $h = \qty{120}{\um}$ is much smaller than its \qty{600}{\um}-width.
%MeAsp is an amino acid. Its radius can be estimated from aspartate's radius as \qty{4.04}{\Angst}~\cite{Liu2020}. Casamino acids are a mixture of amino acids. To roughly estimate its Stokes radius, we suppose it is an equimolar mix of all \num{20} essential amino acids. The average Stokes radius for molecules in such a mix is \qty{4.19}{\Angst}. The diffusion coefficients yielded by these radii are \qty{524}{\square\um\per\s} and \qty{543}{\square\um\per\s}. In both cases, we notice their diffusivity is higher than fluorescein's one. Hence, in equal conditions, their concentration profiles will reach a steady state more rapidly than fluorescein's.

\section{Effect of the chemoattractant concentration and gradient on the swimming velocity and motility}
\label{sec:mot}

To investigate the impact of chemoattractants on bacterial motility, we analysed each video to quantify the bacterial swimming velocity $v_s$ and the trajectory correlation time $\tau$. The same analysis was performed on experiments without chemoattractants to determine their influence on bacterial motility. 

\begin{figure}[htpb]
    \centering
    \includegraphics[width = \linewidth]{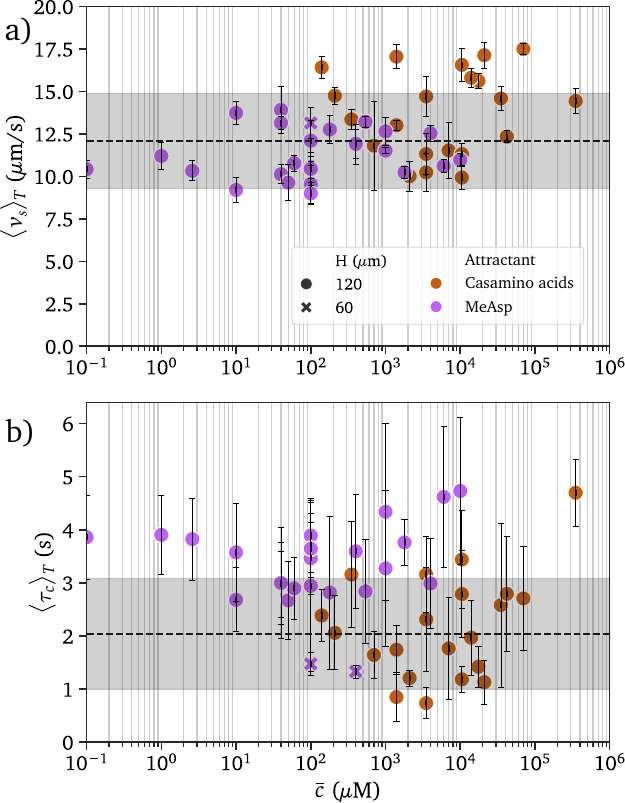}
    \caption{\textbf{Bacterial motility characteristics} a) Bacterial swimming velocity $\langle \bar{v}_s \rangle_T$ and b) average correlation time $\langle \tau \rangle_T$ estimated from the velocity correlation function, as a function of chemoattractants concentration. Purple and orange markers correspond respectively to MeAsp and casamino acids. The average is done over all the image sequences recorded during one experiment. Horizontal dashed lines and grey strips: average and range of variation for three experiments without chemoattractant ($c_1 = c_3 = \qty{0}{\mM}$). Vertical bars indicate the second moment of the time distributions.}
    \label{fig:mot}
\end{figure}

Figure~\ref{fig:mot} displays the averaged values of $v_s$ and $\tau$ from seven time-lapse images recorded at $z = h/2$ during each experiment. The vertical bars represent the quadratic deviation over the seven values measured. In the figures, the values obtained for the experiments conducted without a chemoattractant are shown with a dotted line and a grey background. Each symbol denotes a unique batch of bacteria, making measurement variations likely due to differences between batches. 

Table~\ref{tab:dif} gives the values of the swimming velocity and correlation time averaged over all the experiments performed with or without chemoattractants. The diffusion coefficients $\mu$ derived from the equation $\mu = v_s^2 \tau$ are consistent with the range reported in the literature (see, for example, Tab.~2 in~\cite{Lewus_2001} or~\cite{Ahmed_Shimizu_Stocker_2010}). Our results indicate no significant variation in either the correlation time or the swimming velocity with $\bar{c}$. However, a slight dependence on the chemical nature of the attractant is observed: the correlation time is approximately \qty{1}{\s} longer in MeAsp compared to casamino acids or in the absence of a chemoattractant. Concurrently, the swimming velocity is found to be $\qty{2}{\um\per\second}$ higher in casamino acids than in MeAsp or with out chemoattractant conditions. Ultimately, due to the opposing effects, the diffusion coefficient in MeAsp is comparable to that measured in the absence of chemoattractants (see Table~\ref{tab:dif}).

\begin{table}[htpb]
    \begin{center}
        \begin{tabular}{|c|c|c|c|}
            \hline
            Solute & $v_s$ (\unit{\um\per\s}) & $\tau$ (\unit{\s}) & $\mu$ (\unit{\square\um\per\s})\\
            \hline
            \o & \num{12(3)}& \num{2.0(10)} & \num{290(200)}\\
            \hline
            Cas. acids & \num{14(3)} & \num{2.2(10)} & \num{430(200)}\\
            \hline
            MeAsp & \num{11(1)} & \num{3.3(8)} & \num{400(100)}\\
            \hline
        \end{tabular} 
        \caption{Swimming velocity $v_s$, correlation time $\tau$, and diffusion coefficient $\mu$ in the absence of chemoattractant (\o), with casamino acids (Cas. acids) and with $\alpha$-methyl-DL-aspartic (MeAsp). The values are the averages over all time-lapse images taken at half-height, $z = h/2$, in the channel.}
        \label{tab:dif}
    \end{center}
\end{table}

\section{Verification of the shallow gradient assumption}
\label{sec:shal}

\begin{figure}[htpb]
    \centering
    \includegraphics[width = \linewidth]{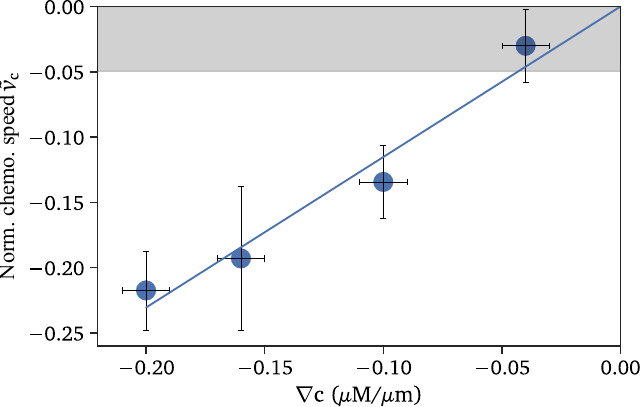}
    \caption{Normalised chemotactic velocity $\langle \tilde v_c(T) \rangle_T$, averaged over all times $T$, as function of $\nabla c$ measured for an average MeAsp concentration of $\bar{c} = \qty{100}{\uM}$. The solid line is a linear fit of the data by $\tilde v_c = \chi \nabla c$ with $\chi = \qty{0.72(5)}{\um\per\uM}$. The vertical bars represent the second moment of the distribution of $\tilde v_c(T)$.}
    \label{fig:gradient}
\end{figure}

A series of experiments was conducted to verify the shallow gradient limit in our studies and to validate the use of Eq.~\eqref{chi(c)}. To achieve this, the values of $c_1$ and $c_3$ were adjusted so that the average concentration $\bar{c}$ remained constant, while the gradient varied. Fig.~\ref{fig:gradient} shows the normalised averaged chemotactic velocity $\tilde{v}_c$ as a function of the imposed gradient. The data points align on a straight line with a slope of $\chi = \qty{0.75}{\um\per\uM}$. Based on this result, we see that the chemotactic susceptibility can be obtained from the measurement of $v_c$ knowing the imposed gradient $\nabla c$ as: $\chi(\bar{c}) = v_c / \nabla c$.

\section{Changes in bacterial profiles observed over time-lapse images}
\label{sec:TimeSeparation}

Eq.~(\ref{eq:ks}) represents the balance between the different fluxes present, specifically the chemotactic and diffusive fluxes, and accounts for flux variations in both space and time. Our method focuses on determining fluxes from analysed tracks generated from 20 seconds of time-lapse images. We assume that the flux remains relatively constant during this 20-second interval. This assumption is equivalent to the time-separation condition used in statistical physics to describe how a system evolves toward equilibrium. The time separation condition implies that correlations between the system’s states at two widely separated times decay. This allows us to describe how microscopic interactions lead to macroscopic properties over time.

\begin{figure}[htpb]
    \centering
    \includegraphics[width = \linewidth]{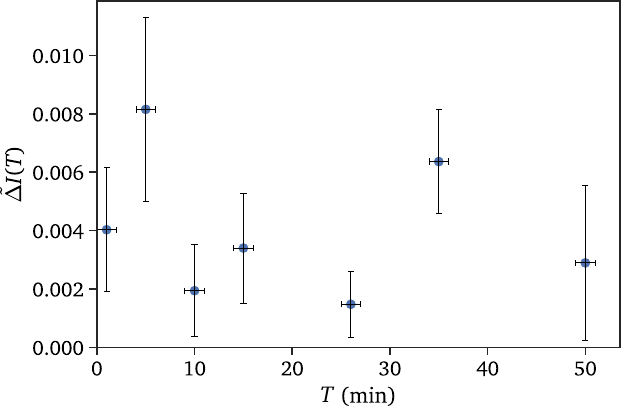}
    \caption{Root-mean-square (RMS) of the difference between the $x$, $t$-averaged light profiles along $y$ at $z = h/2$ measured for the \num{7} image sequences in a single experiment, as a function of time $T$.\\Here, $\tilde{\Delta} I(T) = \left\langle 2 \dfrac{I(T,t > \qty{15}{\s},y) - I(T,t < \qty{5}{\s},y)}{I(T,t < \qty{15}{\s},y) + I(T,t > \qty{5}{\s},y)}\right\rangle_y$.}
    \label{fig:time}
\end{figure}
\newpage

To quantify this point, we analysed the evolution of the bacterial profile over \qty{20}{\s} using the \num{200} frames that compose each time-lapse. First, the intensity of each frame was averaged along the $x$-direction. The first \num{50} profiles were then averaged, and the same operation was performed on the last \num{50} profiles obtained from the final \num{50} frames of the time-lapse. The evolution between the beginning and end of the sequence was quantified using the root-mean-square (RMS) of the difference between the two normalised profiles. This value was measured for the \num{7} image sequences in a single experiment. Fig.~\ref{fig:time} shows the measured value as a function of $T$. This quantity remains consistently below \qty{1}{\percent}, demonstrating that the bacterial profile changes minimally between the start and end of a film. This observation holds throughout the experiment, even as the bacterial profile transitions from a constant distribution to an exponential one. This measurement validates the time-separation condition and confirms that the system’s macroscopic evolution can indeed be described from microscopic results.

\end{document}